\RequirePackage[2020-02-02]{latexrelease}
\documentclass[twocolumn,english,aps,prb,10pt,superscriptaddress]{revtex4}
\usepackage[T1]{fontenc}
\usepackage{amsmath,graphicx,amssymb,epsfig,dsfont,mathrsfs,color,amsbsy}

\begin{document}

\title{Many-Body Floquet Theory for Radiative Heat Transfer in Time-Modulated Systems}

\author{R. Messina}
\affiliation{Laboratoire Charles Fabry, UMR 8501, Institut d'Optique, CNRS, Universit\'{e} Paris-Saclay, 2 Avenue Augustin Fresnel, 91127 Palaiseau Cedex, France}
\author{P. Ben-Abdallah}
\email{pba@institutoptique.fr} 
\affiliation{Laboratoire Charles Fabry, UMR 8501, Institut d'Optique, CNRS, Universit\'{e} Paris-Saclay, 2 Avenue Augustin Fresnel, 91127 Palaiseau Cedex, France}

\date{\today}


\begin{abstract}
We develop a general theory of radiative heat exchange between dipoles with time-modulated optical properties. This framework extends fluctuational electrodynamics beyond equilibrium by incorporating nonstationary correlations and memory effects induced by temporal modulation. Closed-form expressions for the heat currents in modulated many-body systems are obtained, together with a generalized Landauer-like formulation of the pairwise exchanges, where the transmission coefficient accounts for all inelastic frequency-conversion channels. Near-resonant modulation redistributes and amplifies thermal fluctuations across Floquet sidebands, acting as a parametric amplifier of thermal radiation and enabling active, frequency-selective control of nanoscale heat transfer.

\end{abstract}

\maketitle

\section{Introduction}

Controlling thermal radiation through temporal modulation of optical properties offers new opportunities to overcome the fundamental limits imposed by time-invariance and reciprocity. In a pioneering work~\cite{Liberal}, Vázquez-Lozano and Liberal developed a quantum framework for thermal emission in time-modulated solids, predicting radiation beyond the blackbody limit and even vacuum amplification. Subsequently, Yu and Fan~\cite{Fan1} demonstrated that temporal modulation of optical properties in surface-polariton systems enables control of spatial coherence, paving the way for coherent thermal emission. More recently Wang and collaborators studied  the behavior of single emitters~\cite{Wang1,Wang2} and heat exchanges in two-body systems~\cite{Wang3,Wang4}. In the slow-modulation regime, Latella et al.~\cite{Latella,Ma} analyzed heat shuttling driven by periodic modulation of temperature or chemical potential, revealing enhanced transfer and negative differential thermal resistance. Messina and Ben-Abdallah~\cite{Messina} proposed radiative heat pumping via periodic modulation, while Ben-Abdallah and Biehs~\cite{Biehs} identified Berry-like geometric phases as a mechanism for nonreciprocal thermal relaxation. Using Floquet theory, Fernández-Alcázar et al.~\cite{Alcazar} showed that temporal modulation can induce directional heat flow between reservoirs at equal temperatures. Beyond the adiabatic limit, Fan and Yu~\cite{Fan2} introduced a time-dependent dyadic Green’s function formalism that enables active control of radiative heat transfer—including enhancement, suppression, and reversal—and later proposed photonic refrigeration through thermal photon up-conversion~\cite{Fan3}. Despite these advances, a comprehensive framework for describing strongly coupled, many-body, and far-from-equilibrium modulated systems is still lacking. In this work, we develop such a framework by extending fluctuational electrodynamics to time-modulated many-body systems and generalizing the fluctuation–dissipation theorem beyond equilibrium.

\section{Coupled dipole moments in time-modulated many-body systems}
We consider a set of $N$ finite-size thermal emitters at temperatures $T_i$ ($i=1,2,\dots,N$) immersed in a thermal bath at temperature $T_\mathrm{b}$. The objects are assumed small compared to their thermal wavelength, so they can be described as electric dipoles. We further assume that their polarizabilities can be actively modulated in time. The spectral component of each dipole moment can be decomposed as
\begin{equation}
	\mathbf{p}_{i}(\omega)=\mathbf{p}^{\mathrm{(fl)}}_{i}(\omega)+ \mathbf{p}^{\mathrm{(ind)}}_{i}(\omega)+\delta\mathbf{p}^{\mathrm{(ind)}}_{i}(\omega),
	\label{Eq:dipole}
\end{equation}
where $\mathbf{p}^{\mathrm{(fl)}}_{i}(\omega)$ is the fluctuating part of dipolar moment due to thermal fluctuations, whereas $\mathbf{p}^{\mathrm{(ind)}}_{i}(\omega)$ and $\delta\mathbf{p}^{\mathrm{(ind)}}_i(\omega)$ correspond to the induced parts of the dipolar moment, due to the surrounding emitters and thermal bath, induced by the polarizability $\alpha_i$ and its time-modulated part $\delta\alpha_i$, respectively. The first two terms correspond to the usual terms in the fluctuational-electrodynamics theory of many-body systems~\cite{pba2011,messina2013,biehs1}. On the other hand, the third term results from the external modulation of the polarizabilities. This driven term is related to the local electric field by an expression of the following form~\cite{Monticone,Mirmoosa}
\begin{equation}
\delta\mathbf{p}^{\mathrm{(ind)}}_{i}(\omega)=\varepsilon_0\int d\Omega\, \delta\alpha_i(\Omega,\omega-\Omega) \mathbf{E}^{\mathrm{(loc)}}_{i}(\Omega) ,
 \label{Eq:modulation dipole}
\end{equation}
where $\delta\alpha_i$ is polarizability kernel which depends on the external driving of optical properties. Here, $\mathbf{E}^{\mathrm{(loc)}}_{i}(\Omega)$ is the local field taking into account the self-radiated part by the particle under the external modulation, which reads 
\begin{equation}
	\mathbf{E}^{\mathrm{(loc)}}_{i}(\Omega)=\mathbf{E}_{\mathrm{b}i}(\Omega)+\mu_{0}\Omega^{2}\sum_{j=1}^N\mathds{G}_0(\mathbf{r}_i,\mathbf{r}_j,\Omega)\mathbf{p}_{j}(\Omega),
	\label{Eq:loc field}
\end{equation}
where $\mathbf{E}_{\mathrm{b}i}(\Omega)$ is the field at the location of dipole $i$ coming from the thermal bath while the second term denotes the field resulting from all dipoles, including the $i$-th dipole and $\mathds{G}_0(\mathbf{r}_i,\mathbf{r}_j,\Omega)$ is the propagator in vacuum. It is worthwhile to note that unlike $\mathbf{\delta p}_{i}(\omega)$, $\mathbf{p}^{\mathrm{(ind)}}_{i}(\omega)$ is defined from the surrounding dipoles and from the field radiated by the external bath, i.e.
\begin{equation}\begin{split}
	\mathbf{p}^{\mathrm{(ind)}}_{i}(\omega)&=\alpha_i(\omega) \Bigl(\frac{\omega}{c}\Bigr)^2 \sum_{j\neq i}\mathds{G}_0(\mathbf{r}_i,\mathbf{r}_j,\omega)\mathbf{p}_{j}(\omega)\\
	&\,+\varepsilon_0\alpha_{i}(\omega)\mathbf{E}_{\mathrm{b}i}(\omega).
	\label{Eq:induced part}
\end{split}\end{equation}
It is important to note that the spectral convolution in Eq.~\eqref{Eq:modulation dipole} introduces a memory-like effect. Indeed, the dipolar moment induced at the frequency $\omega$ by the modulation of optical properties depends on the local field undergone by the particle over the full spectral range. Combining Eqs.~\eqref{Eq:dipole}, \eqref{Eq:modulation dipole} and \eqref{Eq:induced part}, we see that the the spectral components of dipolar moments are solution of the system of Fredholm's integral equation
\begin{equation}\begin{split}
		&\mathbf{p}_{i}(\omega)-\alpha_{i}(\omega) \Bigl(\frac{\omega}{c}\Bigr)^2\sum_{j\neq i}\mathds{G}_0(\mathbf{r}_i,\mathbf{r}_j,\omega)\mathbf{p}_{j}(\omega)\\
		&\,-\int d\Omega\,\mathbf{\delta \alpha}_i(\Omega,\omega-\Omega)\Bigl(\frac{\Omega}{c}\Bigr)^2\sum_{j=1}^N\mathds{G}_0(\mathbf{r}_i,\mathbf{r}_j,\Omega)\mathbf{p}_{j}(\Omega) \\
		&=\mathbf{p}^{\mathrm{(fl)}}_{i}(\omega)+\varepsilon_0\alpha_{i}(\omega)\mathbf{E}_{\mathrm{b}i}(\omega)\\
		&\,+\varepsilon_0\int d\Omega\,\mathbf{\delta \alpha}_i(\Omega,\omega-\Omega)\mathbf{E}_{\mathrm{b}i}(\Omega).
		\label{Eq:dipole2}
\end{split}\end{equation}
This equation forms the foundation of radiative heat transfer in time-modulated many-body systems. It can be written in a compact block-matrix form and solved perturbatively to arbitrary order in $\delta\alpha$, allowing calculation of dipole correlations and the resulting thermal emission spectrum. By distinguishing in the integral term the contribution coming from the surrounding dipoles (i.e. $\mathbf{r}_i\neq \mathbf{r}_j$) from the self contribution of each particle and by rewriting the previous equation into a block integral equation we get
\begin{equation}\begin{split}
	&\mathds{A}(\omega)\mathbf{p}(\omega)-\int d\Omega\,\mathds{K}(\Omega,\omega-\Omega)\mathbf{p}(\Omega)=\mathbf{S}(\omega)+\delta\mathbf{S}(\omega),
\end{split}	\label{Eq:dipole3}\end{equation}
where we have set the $(3\times N)$-dimension dipolar moment $\mathbf{p}(\omega)=(\mathbf{p}_{1}(\omega),\dots,\mathbf{p}_{N}(\omega))^T$ and the source terms $\mathbf{S}(\omega)=(\varepsilon_0\alpha_1(\omega) \mathbf{E}_{\mathrm{b}1}(\omega)+\mathbf{p}^{\mathrm{(fl)}}_{1}(\omega),\dots,\varepsilon_0\alpha_N(\omega) \mathbf{E}_{\mathrm{b}N}(\omega)+\mathbf{p}^{\mathrm{(fl)}}_{N}(\omega))^T$ and $\delta\mathbf{S}(\omega)=\varepsilon_0\bigl(\int d\Omega\,\mathbf{\delta \alpha}_1(\Omega,\omega-\Omega)\mathbf{E}_{\mathrm{b}1}(\Omega),\dots,\int d\Omega\,\mathbf{\delta \alpha}_N(\Omega,\omega-\Omega)\mathbf{E}_{\mathrm{b}N}(\Omega)\bigr)^T$, while the block matrices $\mathds{A}$ and $\mathds{K}$ are defined as follows
\begin{equation}\begin{split}
	\mathds{A}_{ij}(\omega)&= \delta_{ij}\mathds{1} +(1-\delta_{ij})\alpha_{i}(\omega) \Bigl(\frac{\omega}{c}\Bigr)^2 \mathds{G}_0(\mathbf{r}_i,\mathbf{r}_j,\omega),\\
	\mathds{K}_{ij}(\omega,\omega')&=\delta\alpha_i(\omega,\omega')\frac{\omega^2}{c^2}\Bigl[ \delta_{ij}\mathds{G}_0(\mathbf{r}_i,\mathbf{r}_j\to\mathbf{r}_i,\omega)\\
		&\,+(1-\delta_{ij})\mathds{G}_0(\mathbf{r}_i,\mathbf{r}_j,\omega)\Bigr],
	\end{split}
	\label{block}\end{equation}
where, according to Ref.~\onlinecite{Lakhtakia}, the propagator in vacuum asymptotically reads
	\begin{equation}
		\mathds{G}_0(\mathbf{r},\mathbf{r}'\to\mathbf{r},\omega)=G_0(\omega)\mathds{1}=\Bigl(-\frac{c^2}{3\omega^2V}+i\frac{\omega}{6\pi c}\Bigr)\mathds{1},\end{equation}
where $V$ is the particle volume. 

\section{Perturbative expansion of dipolar moments}
The dipolar moments can be calculated in a perturbative way using the ansatz
\begin{equation}
	\mathbf{p}(\omega)= \sum_{n=0}^{\infty}\mathbf{p}^{(n)}(\omega),
	\label{Eq:perturbation}
\end{equation}
where $\mathbf{p}^{(n)}(\omega)$ is of order $n$ in the perturbation $\delta\alpha$. By inserting this expression into Eq.~\eqref{Eq:dipole3} we get for the unmodulated system $\mathbf{p}^{(0)}(\omega)=\mathds{A}^{-1}(\omega)\mathbf{S}(\omega)$, whereas for order 1 and any $n>1$ we have
\begin{equation}\begin{split}
	\mathbf{p}^{(1)}(\omega)&= 
 \mathds{A}^{-1}(\omega)\Biggl[\int d\Omega\,\mathds{K}(\Omega,\omega-\Omega)\mathbf{p}^{(0)}(\Omega)+\delta\mathbf{S}(\omega)\Biggr],\\
	\mathbf{p}^{(n)}(\omega)&= 	\mathds{A}^{-1}(\omega)\int d\Omega\,\mathds{K}(\Omega,\omega-\Omega)\mathbf{p}^{(n-1)}(\Omega).\end{split}	\label{recursion2}\end{equation}
These recursive relations allow us to write a general expression relating the $n^\mathrm{th}$ order ot the dipolar moment to the dipolar moments ${\mathbf{p}}^{(0)}$ of the unmodulated system and to the field $\mathbf{E}_{\mathrm{b}}$ associated with the external bath by the following expression
\begin{equation}
		\mathbf{p}^{(n)}(\omega)= 	\mathds{O}^{(n)}(\omega)\mathbf{p}^{(0)}+\mathds{O}^{(n-1)}(\omega)\delta\mathbf{S},\label{recursion3}\end{equation}
where we have defined the integral operator
\begin{equation}
\begin{split}
	{\mathds{O}}^{(n)}(\omega)\mathbf{f} &\equiv\int d\Omega_1\,\mathds{P}(\omega,\Omega_1)\int d\Omega_2\,\mathds{P}(\Omega_1,\Omega_2)\times\dots\\
&\,\times\int d\Omega_n\,\mathds{P}(\Omega_{n-1},\Omega_n)\mathbf{f}(\Omega_n),
\end{split}
\end{equation}
being $\mathds{P}(\omega,\omega')=\mathds{A}^{-1}(\omega)\mathds{K}(\omega',\omega-\omega')$. The cross-correlation functions of dipolar moments in the modulated system can thus also be formally expressed as perturbative expansions over the modulation order, generalizing the standard correlations in a static system. In particular, the background field and intrinsic dipole fluctuations obey the usual fluctuation-dissipation relations~\cite{biehs1,messina2013,pba2024}. 

\begin{figure}
	\centering
	\includegraphics[width=0.48\textwidth]{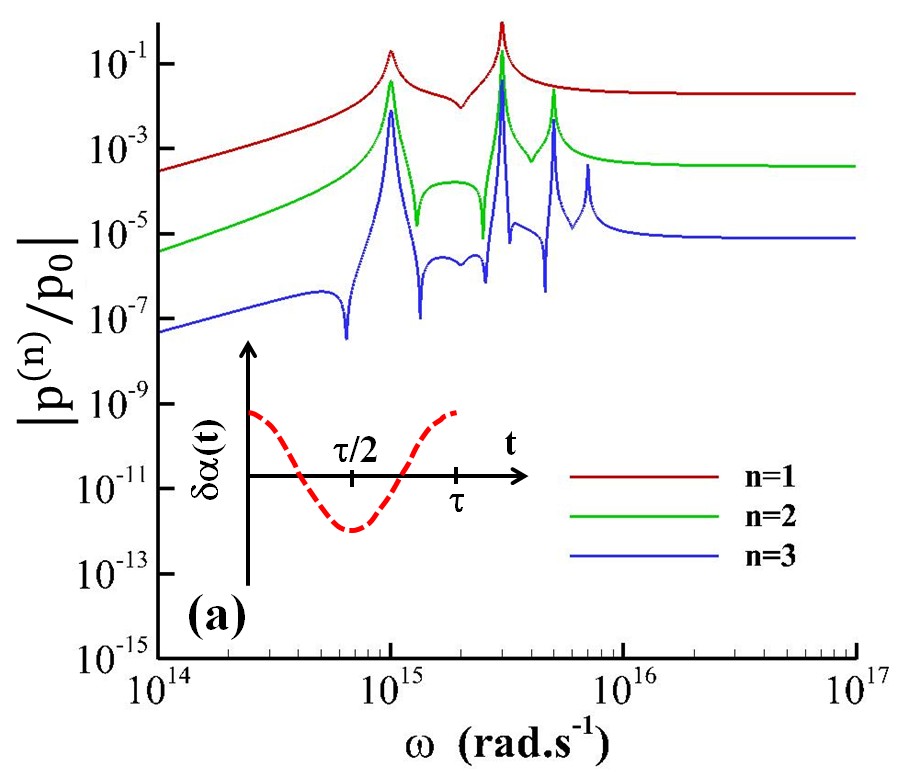}
	\includegraphics[width=0.48\textwidth]{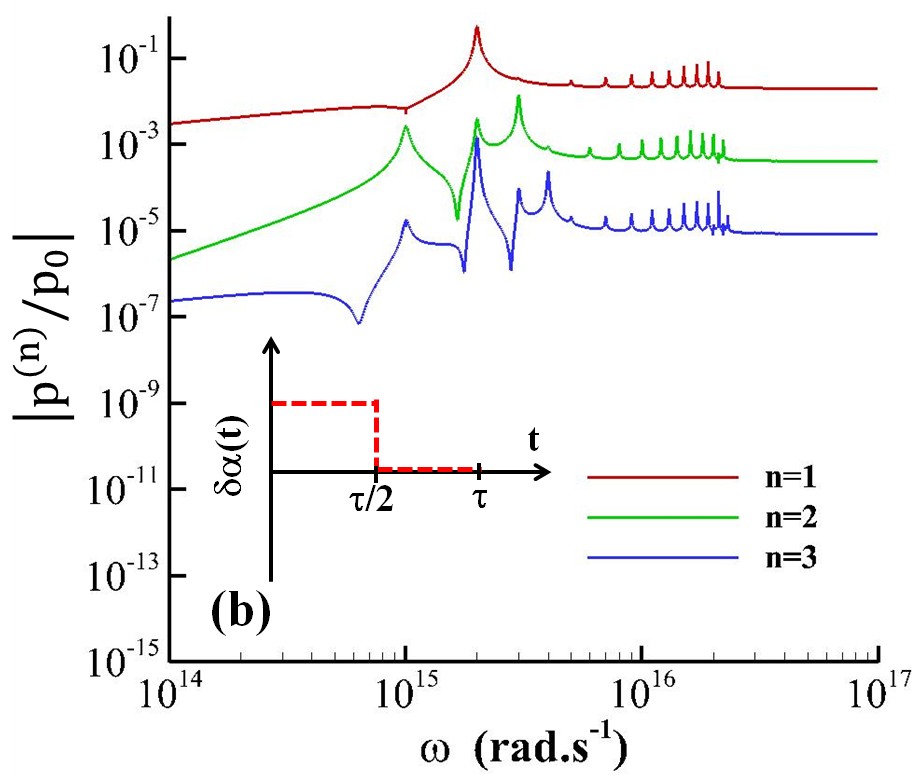}
	\caption{Dipolar corrections of order $n$ in the scalar approximation with (a) a sinusoidal variation of $\delta\alpha$ of period $\tau=2\pi/\Omega$ and with (b) a rectangular variation $\delta\alpha(\omega,t)$ with the same period ($\delta\alpha(\omega,t)$ is equal to one when $\tau\in[0,\tau/2]$ and zero elsewhere and it is decomposed in 20 Fourier modes). Here we take $\omega_0=10^{15}\,\mathrm{rad}\cdot\mathrm{s}^{-1}$, $\gamma=5\cdot10^{13}\,\mathrm{s}^{-1}$, $\alpha_0=0.001$, $\tilde{p}=0.01$ and the temperature is $T=300\,$K. Here, the modulation frequency coincides with the resonance frequency of the Lorentzian (i.e. $\Omega=\omega_0$). Inset: temporal variation of $\delta\alpha$.}
\label{Fig_scalar}\end{figure} 

\section{Dipolar corrections in the scalar approximation}
To describe how the polarizability modulations impact the dipolar moments, we consider below a simple scenario in which $\delta\alpha$ is a $\tau$-periodic function ($\tau=2\pi/\Omega$) and thus can be written as 
\begin{equation}
 \delta\alpha(\omega,\omega') =\delta\hat{\alpha}(\omega)\underset{n}{\sum}\beta_n \delta(\omega'+n\Omega),
 \label{kernel}
\end{equation}
where $\beta_n$ are the Fourier components of the temporal variation of $\delta\alpha(\omega,t)$.

We start from the case of a single scalar dipole. By employing Eq.~\eqref{recursion3} and neglecting the external bath, the dipolar correction of order $m\geq 1$ reads
\begin{equation}
p^{(m)}(\omega) = \sum_{\substack{n_1, \dots, n_m }}U_{n_1,\dots,n_m}(\omega,\Omega)
p_0 \Bigl( \omega + \sum_{k=1}^{m} n_k \, \Omega \Bigr) ,
 \label{corrections2}
\end{equation}
with
\begin{equation}
\begin{aligned}
U_{n_1,\dots,n_m}(\omega,\Omega)=\hat{p}(\omega) \times\prod_{k=1}^{m} \beta_{n_k}\prod_{k=2}^{m} \hat{p} \Bigl( \omega + \underset{j=2}{\overset{k}{\sum}}n_j\, \Omega\Bigr).
\end{aligned}
 \label{corrections3}
\end{equation}
As far as $\hat{p} $ is concerned, it is chosen, for the sake of clarity, proportional to $\delta\hat{\alpha}$ (i.e. $\hat{p}=\tilde{p}\frac{\omega^2}{\omega_0^2-\omega^2-i\gamma\omega}$).
In Fig.~\ref{Fig_scalar} we show the evolution of dipolar corrections in the specific case where the spectral components of $p_0$ reads
\begin{equation}
 p_0(\omega) =\sqrt{S(\omega)} \xi(\omega).
 \label{p0}
\end{equation}
In this expression $\xi$ is a complex white noise and $S$ is the power spectral density of thermal fluctuations at a given temperature $T$ defined as
\begin{equation}
 S(\omega) =4\mathscr{C}(\omega,T)\mathrm{Im}[\alpha(\omega)]
 \label{thermal_fluc}
\end{equation}
where $\mathscr{C}$ denotes the occupation function at a given temperature
\begin{equation}
\mathscr{C}(\omega,T) =\frac{\hbar}{2}\coth\frac{\hbar\omega}{2k_B T}.
 \label{occuption}
\end{equation}
and $\mathrm{Im}[\alpha]$ is the imaginary part of a Lorentzian polarizability i.e.
\begin{equation}
\mathrm{Im}[\alpha(\omega)] =\alpha_0\frac{\omega^2_0\gamma \omega}{(\omega^2_0-\omega^2)^2+(\gamma\omega)^2}.
 \label{im_polar}
\end{equation}
When the polarizability is periodically modulated, Fig.~\ref{Fig_scalar} shows that the spectrum of dipole moments is no longer confined to the single resonance frequency of the static Lorentzian response. Instead, it is redistributed into a series of sidebands at frequencies $\omega+n\Omega$, where $\Omega$ denotes the modulation frequency. As a result, the spectral profile evolves into a comb-like structure, with amplitudes determined by the strengths of the modulation harmonics. In agreement with Eq. (\ref{corrections2}), the spectra associated with higher perturbative orders exhibit a progressively broader frequency spread compared to the lower-order components.

\section{Radiative heat exchanges in time-modulated many-body systems}
Going back to the general vector approach, it is possible to determine the heat dissipated by each object in a general time-modulated dipolar system. The mean power received by the $i^{\rm th}$ dipole is defined as~\cite{messina2013}
\begin{equation}
	\mathcal{P}_{i} = 2\,{\rm Im}\int_{0}^{\infty}\!\frac{{\rm d}\omega}{2\pi}\, \omega\! \int_{0}^{\infty}\frac{{\rm d}\omega'}{2\pi}\langle\mathbf{p}_i(\omega)\cdot \mathbf{E}_{i}^*(\omega')\rangle\;e^{-i(\omega-\omega')t}. 
\label{Eq:powerNdipoldef}
\end{equation}
This expression allows to compute the radiative heat flux between two dipoles in a time-modulated system, which can be expressed in a generalized Landauer form, extending the standard fluctuational electrodynamics framework. 

\subsection{Cross-correlation functions of dipolar moments}

The two-frequency correlation matrix is defined as
\begin{equation}
\mathds{C}_{ij}(\omega,\omega') \equiv \langle \mathbf{p}_i(\omega) \otimes \mathbf{p}_j^*(\omega') \rangle,
\end{equation}
where $i,j=1,\dots,N$ label the dipoles. In a non-modulated (stationary) system, this reduces to
\begin{equation}
\mathds{C}_{ij}(\omega,\omega') 
= \sum_k \mathds{A}^{-1}_{ik}(\omega) 
\langle \mathbf{S}_k(\omega) \otimes \mathbf{S}^*_k(\omega') \rangle
\mathds{A}^{-1\dagger}_{jk}(\omega'),
\end{equation}
where $\mathbf{S}_k$ contains the fluctuating sources: intrinsic dipole fluctuations $\mathbf{p}^{\mathrm{(fl)}}$ and background field $\mathbf{E}_\mathrm{b}$. 
The correlations of these sources are obtained using the standard FDT:
\begin{equation}\begin{split}
\langle E_{\mathrm{b},i}(\omega) E_{\mathrm{b},j}^*(\omega') \rangle &= 2 \omega^2 \mu_0 \hbar \left(n_b + \frac12\right) \frac{\mathds{G}_{ij}-\mathds{G}_{ji}^\dagger}{2i},\\
\langle p^{\mathrm{(fl)}}_i(\omega) \otimes p^{\mathrm{(fl)}*}_j(\omega') \rangle &= 4 \pi \varepsilon_0 \hbar \, \delta_{ij} \, \delta(\omega-\omega') \left(n_i + \frac12 \right)\\
&\,\times\mathrm{Im}[\alpha_i(\omega)].
\end{split}\end{equation}

\subsection{Time-modulated systems and Floquet decomposition}

For a system modulated periodically at frequency $\Omega$, correlations acquire a sideband structure:
\begin{equation}
\mathds{C}_{ij}(\omega,\omega') = \sum_n \mathds{C}_{ij}^{(n)}(\omega)\, \delta(\omega' - \omega - n \Omega),
\end{equation}
where $\mathds{C}_{ij}^{(n)}(\omega)$ is the Floquet correlation matrix corresponding to the $n$-th sideband. The dipole dynamics can be expressed in operator form
\begin{equation}
\mathds{N}\cdot\mathbf{p}\equiv\mathds{A}(\omega) \mathbf{p}(\omega) - \int d\Omega \, \mathds{K}(\Omega,\omega-\Omega) \mathbf{p}(\Omega)= \mathbf{S}(\omega),
\end{equation}
where $\mathds{K}$ encodes frequency mixing due to modulation. The inverse operator admits a Floquet decomposition
\begin{equation}
\mathds{N}^{-1}(\omega,\omega') = \sum_n \alpha_n(\omega) \delta(\omega' - \omega - n \Omega),
\end{equation}
where $\alpha_n(\omega)$ is the $n$-th Floquet component of the generalized polarizability.
\subsection{Generalized fluctuation--dissipation theorem} 
We start from the most general linear response relation between the induced dipole moment $\mathbf{p}(t)$ and an applied electric field $\mathbf{E}(t)$ in a nonstationary medium, \begin{equation} 
\mathbf{p}(t) = \varepsilon_0 \int_{-\infty}^{\infty}\! dt'\alpha(t,t')\mathbf{E}(t'), \label{eq:alpha_time} 
\end{equation} 
where $\alpha(t,t')$ is the (causal) two-time polarizability kernel, including both the static and the time-varying contributions [i.e. $\alpha(\omega,\omega')\equiv\alpha(\omega)\delta(\omega-\omega')+\delta\alpha(\omega,\omega')$]. In a time-translation invariant (stationary) system $\alpha(t,t')=\alpha(t-t')$ and one works with a single-frequency susceptibility; here the kernel depends on both times. Assume the system is driven by a periodic modulation with angular frequency $\Omega$, so that
\begin{equation}
\alpha(t+\mathcal{T},t'+\mathcal{T})=\alpha(t,t'),\qquad \mathcal{T}=2\pi/\Omega.
\end{equation}
It is then convenient to work in the double Fourier (frequency) domain and to exploit the Floquet structure. Define the mixed frequency representation 
\begin{equation} 
\alpha(\omega,\omega') \equiv \int dt\int dt' e^{i\omega t} e^{i\omega' t'} \alpha(t,t'). 
\end{equation} 
Periodic modulation implies that $\alpha(\omega,\omega')$ is nonzero only when $\omega'-\omega$ is an integer multiple of $\Omega$. Equivalently, one can write the Floquet (sideband) decomposition
 \begin{equation}
 \alpha(\omega,\omega')= \sum_{n\in\mathbb{Z}} \alpha_n(\omega) \delta\!\big(\omega' - \omega - n\Omega\big). 
\label{eq:alpha_floquet} 
\end{equation} 
The matrix $\alpha_n(\omega)$ is the $n$-th Floquet component of the generalized polarizability and couples the field at frequency $\omega+n\Omega$ to the response at frequency $\omega$. We now state the fluctuation–dissipation relation. In the stationary (equilibrium) quantum case the Kubo formula relates the symmetrized correlation of an observable to the imaginary part of the corresponding susceptibility. For the dipole fluctuations $\mathbf{p}^{\mathrm{(fl)}}(t)$ in thermal equilibrium at temperature $T$ the (double-frequency) quantum FDT generalizes to the nonstationary, periodically driven situation as follows: the two-frequency correlator of intrinsic dipole fluctuations is proportional to the imaginary (dissipative) part of the two-frequency polarizability kernel, 
\begin{equation} 
\big\langle \mathbf{p}^\mathrm{(fl)}(\omega)\otimes\mathbf{p}^{\mathrm{(fl)}*}(\omega')\big\rangle \;=\; 4\pi\varepsilon_0\hbar\; \mathcal{N}(\omega)\; \mathrm{Im}\big[ \alpha(\omega,\omega') \big], \label{eq:FDT_general_kernel} 
\end{equation} where $\mathrm{Im}[\cdot]$ denotes the anti-Hermitian part with respect to the frequency arguments (i.e.\ the dissipative part), and $\mathcal{N}(\omega)$ is the thermal occupation weight. In the bosonic equilibrium case one can use 
\begin{equation} \mathcal{N}(\omega) = n(\omega)+\tfrac12, 
\end{equation} so that the prefactor recovers the usual $(n+1/2)$ quantum form. Using the Floquet decomposition \eqref{eq:alpha_floquet} the imaginary part of the kernel has the same sideband structure,
\begin{equation}
	 \mathrm{Im}\big[\alpha(\omega,\omega')\big] = \sum_n \mathrm{Im}\big[\alpha_n(\omega)\big] \delta(\omega' - \omega - n\Omega),
\end{equation}
hence Eq.~\eqref{eq:FDT_general_kernel} becomes
 \begin{equation}
\begin{split}
\big\langle \mathbf{p}^\mathrm{(fl)}(\omega)\otimes\mathbf{p}^{\mathrm{(fl)}*}(\omega')\big\rangle 
&= 4\pi\varepsilon_0\hbar \sum_{n\in\mathbb{Z}} \delta(\omega'-\omega-n\Omega) \\
&\quad \times \big[n(\omega)+\tfrac12\big]\; \mathrm{Im}\big[\alpha_n(\omega)\big].
\end{split}
\label{eq:FDT_Floquet}
\end{equation}
This expresssion is the natural Floquet generalization of the stationary FDT. It states that fluctuations at frequency $\omega$ are correlated with fluctuations at $\omega'=\omega+n\Omega$, i.e.\ the modulation induces cross-frequency correlations (inelastic channels).

\subsection{Floquet correlation matrix from generalized FDT}

In the Floquet basis, the dipole moment reads
\begin{equation}
\mathbf{p}_i(\omega) = \varepsilon_0\sum_n \alpha_n^{(i)}(\omega)\, \mathbf{E}_i^{\rm loc}(\omega-n\Omega) + \mathbf{p}_i^\mathrm{(fl)}(\omega),
\end{equation}
where the local field propagates intrinsic dipole fluctuations
\begin{equation}
\mathbf{E}_i^{\rm loc}(\omega) = \mu_0\omega^2\sum_j \mathds{G}_{ij}(\omega)\, \mathbf{p}_j^\mathrm{(fl)}(\omega).
\end{equation}
The two-frequency correlation is
\begin{equation}
\begin{split}
\langle \mathbf{p}_i(\omega) &\otimes \mathbf{p}_j^*(\omega') \rangle \; = \; 
\varepsilon_0^2 \sum_{n,m} \alpha_n^{(i)}(\omega) \\
&\quad \times \langle \mathbf{E}_i^{\rm (loc)}(\omega-n\Omega) 
\otimes \mathbf{E}_j^{\mathrm{(loc)}*}(\omega'-m\Omega) \rangle \\
&\quad \times \alpha_m^{(j)\dagger}(\omega') 
+ \langle \mathbf{p}_i^\mathrm{(fl)}(\omega) \otimes \mathbf{p}_j^{\mathrm{(fl)}*}(\omega') \rangle .
\end{split}
\end{equation}
Now, we use the generalized FDT (see section above) which gives the intrinsic dipole correlation in the presence of time modulation
\begin{equation}
\begin{split}
\langle \mathbf{p}_i^\mathrm{(fl)}(\omega) &\otimes \mathbf{p}_j^{\mathrm{(fl)}*}(\omega') \rangle \; = \; 
4\pi \varepsilon_0 \hbar \sum_{n\in\mathbb{Z}} \delta(\omega'-\omega-n\Omega) \\
&\quad \times [n_i(\omega)+1/2] \, \mathrm{Im}[\alpha_{i,n}(\omega)] \, \delta_{ij}.
\end{split}
\end{equation}
Substituting this into the local-field correlator and collecting terms produces the Floquet correlation matrix
\begin{equation}
\begin{split}
\mathds{C}_{ij}^{(n)}(\omega) &= \sum_{p-q=n} 
\alpha_p^{(i)}(\omega)\Bigl[n_j(\omega+p\Omega)+\frac{1}{2}\Bigr]\\ \, 
&\,\times\mathrm{Im}\bigl[ \alpha_{j,0}(\omega+p\Omega)\bigr]\alpha_q^{(j)\dagger}(\omega).
\end{split}
\end{equation}
showing explicitly how the cross-frequency correlations arise from the generalized FDT.

\subsection{Power exchanged and generalized Landauer formula}

The instantaneous power absorbed by particle $i$ due to the field generated by particle $j$ is
\begin{equation}
P_{i \leftarrow j}(t) = \langle \dot{\mathbf{p}}_i(t) \cdot \mathbf{E}_i(t) \rangle,
\end{equation}
where $\mathbf{E}_i(t)$ is the local field at $i$ due to $j$ and $\dot{\mathbf{p}}_i(t) = \frac{d}{dt}\mathbf{p}_i(t)$. 
Using Fourier transforms,
\begin{equation}
\mathbf{p}_i(t) = \int_{-\infty}^{\infty} d\omega \, \mathbf{p}_i(\omega) e^{-i\omega t}, 
\mathbf{E}_i(t) = \int_{-\infty}^{\infty} d\omega \, \mathbf{E}_i(\omega) e^{-i\omega t},
\end{equation}
the time-averaged power becomes
\begin{equation}
P_{i \leftarrow j} = \int_{0}^{\infty}\frac{{\rm d}\omega'}{2\pi}\, \int_{0}^{\infty}\frac{{\rm d}\omega}{2\pi}(i\omega)\langle\mathbf{p}_i(\omega) \cdot \mathbf{E}_{i}^{*}(\omega')\rangle e^{-i(\omega-\omega')t}. 
\end{equation}
After time-averaging over one period only $\omega=\omega'$ contributes and
\begin{equation}
P_{i \leftarrow j} = \int_0^\infty \frac{d\omega}{2\pi} \, i \omega \, \mathrm{Tr} \langle \mathbf{p}_i(\omega) \otimes \mathbf{E}_i^*(\omega) \rangle.
\end{equation}
The field at $i$ due to dipole $j$, including sidebands from modulation, reads
\begin{equation}
\mathbf{E}_i(\omega) = \sum_n \mathds{G}_{ij}^{(n)}(\omega+n\Omega) \, \mathbf{p}_{j,n}(\omega+n\Omega),
\end{equation}
where $\mathds{G}_{ij}^{(n)}$ is the full Green tensor at the sideband, including all multiple scattering which is obtained from the Dyson equation
\begin{equation}
\mathds{G}^{(n)}(\omega+n\Omega) = \mathds{G}_0(\omega+n\Omega) + \mathds{G}_0(\omega+n\Omega)\, \alpha_0\, \mathds{G}^{(n)}(\omega+n\Omega),
\end{equation}
where \(\alpha_0\) contains the stationary polarizabilities of all particles. 
Using the Floquet correlation matrix $\mathds{C}_{ij}^{(n)}$ the exchanged power reads
\begin{equation}
P_{i \leftarrow j} = \sum_n \int_0^\infty \frac{d\omega}{2\pi} \, i \omega \, 
\mathrm{Tr} \Big[ \mathds{C}_{ij}^{(n)}(\omega) \, \mathds{G}_{ij}^{(n)\dagger}(\omega+n\Omega) \Big].
\end{equation}
This allows us to define the transmission coefficient including all multiple scattering
\begin{equation}
\begin{split}
\mathcal{T}_{ij}^{(n)}(\omega) &= 4 \, 
\mathrm{Tr} \Big[ 
\mathrm{Im}[\alpha_{i,0}(\omega)] \mathds{G}_{ij}^{(n)}(\omega+n\Omega) \\
&\quad \times \mathrm{Im}[\alpha_{j,n}(\omega+n\Omega)]  \mathds{G}_{ij}^{(n)\dagger}(\omega+n\Omega) 
\Big],
\end{split}
\end{equation}
so that the power exchanged from $i$ to $j$ reads
\begin{equation}
P_{i\to j} = \sum_n \int_0^\infty \frac{d\omega}{2\pi} \, \hbar (\omega+n\Omega) \, \big[ n_i(\omega) - n_j(\omega+n\Omega) \big] \, \mathcal{T}_{ij}^{(n)}(\omega).
\end{equation}
This is the generalized Landauer formula for time-modulated media, including all multiple scattering processes at each sideband. This formulation reduces to the standard Landauer coefficient in the absence of modulation, while the reciprocity relation $\mathcal{T}_{ij}^{(n)}(\omega) = \mathcal{T}_{ji}^{(-n)}(\omega+n\Omega)$ guarantees detailed balance.

\begin{figure}
	\centering
	\includegraphics[height=0.45\textwidth,angle=-90]{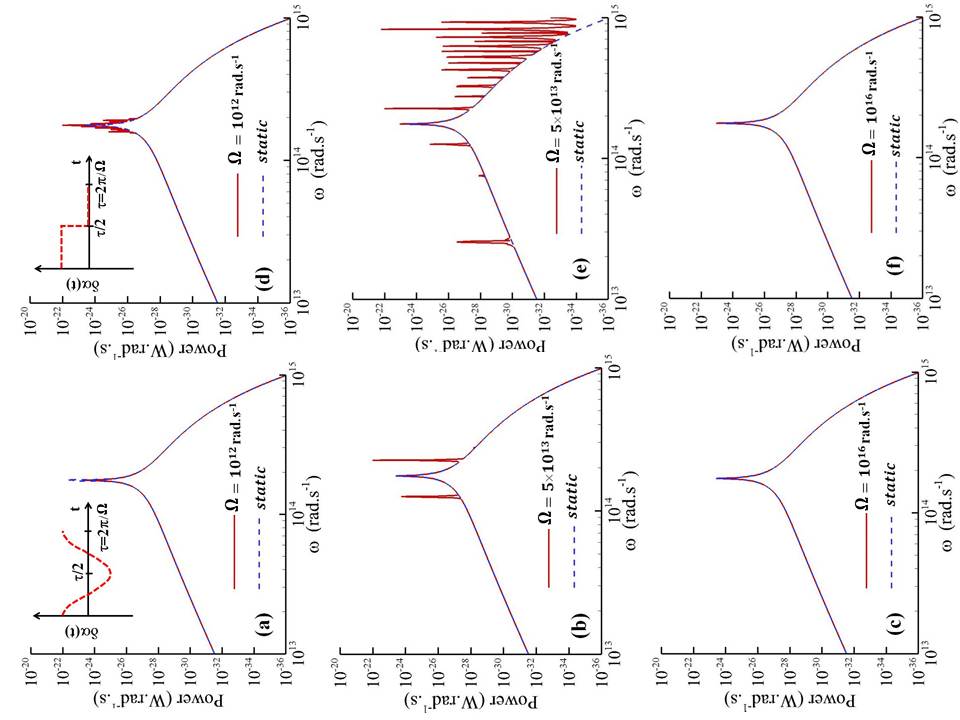}
	\caption{Power spectrum of thermal emission at $T=300\,K$ of a static and modulated dipole associated to a SiC nanoparticle~\cite{Palik} $100\;nm$ radius under a sinusoidal variation (a)-(c) of $\delta\alpha$ of period $\tau=\frac{2\pi}{\Omega}$ and with a rectangular variation (d)-(f) of $\delta\alpha(\omega,t)$ (20 Fourier components are considered). Here we choose an oscillation amplitude of $\delta\alpha_0=0.3 \mathrm{Im}[\alpha]$. Inset: temporal variation of $\delta\alpha$.}
\label{Fig:power_spectrum}\end{figure} 

\section{Thermal emission of a time-modulated nano-source}
To analyze how the time modulation impacts heat flux radiated by an object we consider below the simplest case of nano-source at a given temperature emitting thermal radiation in its background at $0\,$K under a periodic modulation of its polarizability. From the expression (\ref{Eq:loc field}) of local electric field it is direct to see that
\begin{equation}
	\langle\mathbf{p}(\omega)\cdot \mathbf{E}^{\mathrm{(loc)}*}(\omega')\rangle=\mu_0\omega^2G^*_0(\omega')\langle\mathbf{p}(\omega)\cdot \mathbf{p}^*(\omega')\rangle. 
\label{power spectrum}
\end{equation}
Then using the ansatz (\ref{Eq:perturbation}) and the recursive relation \eqref{recursion3} the different perturbation orders read
\begin{equation}
 \mathbf{p}^{(m)}(\omega) =\!\sum_{n_1,\dots,n_m}\! V_{n_1,\dots,n_m}(\omega,\Omega)\mathbf{p}_0\Bigl(\omega+\sum^m_{j=1}n_j\Omega\Bigr),
 \label{emission perturb}
\end{equation}
where 
\begin{equation}\begin{split}
 V_{n_1,\dots,n_m}(\omega,\Omega)&=\prod^m_{k=1}\beta_{n_k}\delta\hat{\alpha}\Bigl(\omega+\sum^k_{j=1}n_j\Omega\Bigr)\\
 &\,\times f\Bigl(\omega+\sum^k_{j=1}n_j\Omega\Bigr),
 \label{perturb_coeff}
\end{split}\end{equation}
with $f(\omega)=\frac{\omega^2}{c^2}G_0(\omega)$. It follows that
\begin{equation}
\begin{split}
\langle&\mathbf{p}^{(m)}(\omega)\cdot\mathbf{p}^{(k)*}(\omega')\rangle\\
&=\sum_{n_1,\dots,n_m \atop n'_1,\dots,n'_k}V_{n'_1,\dots,n'_m}(\omega,\Omega)V^*_{n'_1,\dots,n'_k}(\omega',\Omega)\\
&\,\times\langle\mathbf{p}_0\Bigl(\omega+\sum^m_{j=1}n_j\Omega\Bigr)\cdot \mathbf{p}^*_0\Bigl(\omega+\sum^k_{j=1}n'_j\Omega\Bigr)\rangle.
\label{correlation_correction}
\end{split}
\end{equation}
Then, the time averaged spectrum of perturbative contributions reads
\begin{equation}
\begin{split}
&\mathcal{P}_\mathrm{pert}(\omega) =\frac{12\hbar\omega^3}{c^2}\\
&\,\times\sum'_{n_1,\dots,n_m \atop n'_1,\dots,n'_k} \!\mathrm{Im}\Big[V_{n_1,\dots,n_m}(\omega,\Omega)V^*_{n'_1,\dots,n'_k}(\omega,\Omega)G^*_0(\omega)\Bigr]\\
&\,\times \mathrm{Im}\Bigl[\alpha\Bigl(\omega+\sum^k_{j=1}n_j\Omega\Bigr)\Bigr]n\Bigl(\omega+\sum^k_{j=1}n_j\Omega,T\Bigr).
\label{perturb_contribution}
\end{split}
\end{equation}
Here the sum operator acts over all sets $(n_1,\dots,n_m)$ and $(n'_1,\dots,n'_k)$ satisfying $\sum^m_{j=1}n_j=\sum^k_{j=1}n'_j$ (all other contributions cancel out after time averaging). Finally, the total power spectrum radiated by a time-modulated dipole can be expressed as
\begin{equation}
\mathcal{P}_{\mathrm{total}}(\omega)
= \frac{2\hbar\omega^{4}}{\pi c^{3}}\,
n(\omega,T)\,
\mathrm{Im}\big[\alpha(\omega)\big]
- \mathcal{P}_{\mathrm{pert}}(\omega),
\label{total spectrum}
\end{equation}
or, equivalently, in the generalized Landauer-like form,
\begin{equation}
\mathcal{P}_{\mathrm{total}}(\omega)
= \sum_{n=-\infty}^{\infty}
\hbar\omega\,
n(\omega-n\Omega,T)\,
\mathcal{T}^{(n)}(\omega-n\Omega),
\end{equation}
where $\mathcal{T}^{(n)}(\omega)$ denotes the transmission coefficient (written in term of propagator in vaccum) associated with the inelastic emission or absorption of 
$n$ modulation quanta $\hbar\Omega$ , $\mathcal{T}^{(0)}(\omega)$ representing the stationary contribution of the emitter.
\begin{figure}
	\centering
	\includegraphics[height=0.3\textwidth,angle=0]{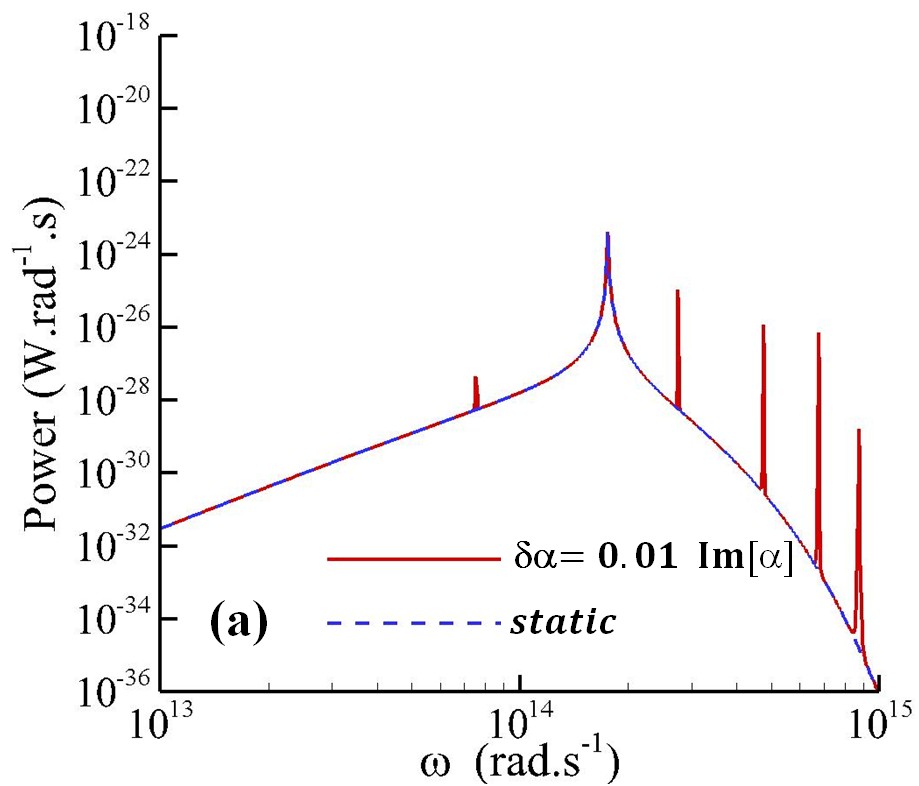}
	\includegraphics[height=0.3\textwidth,angle=0]{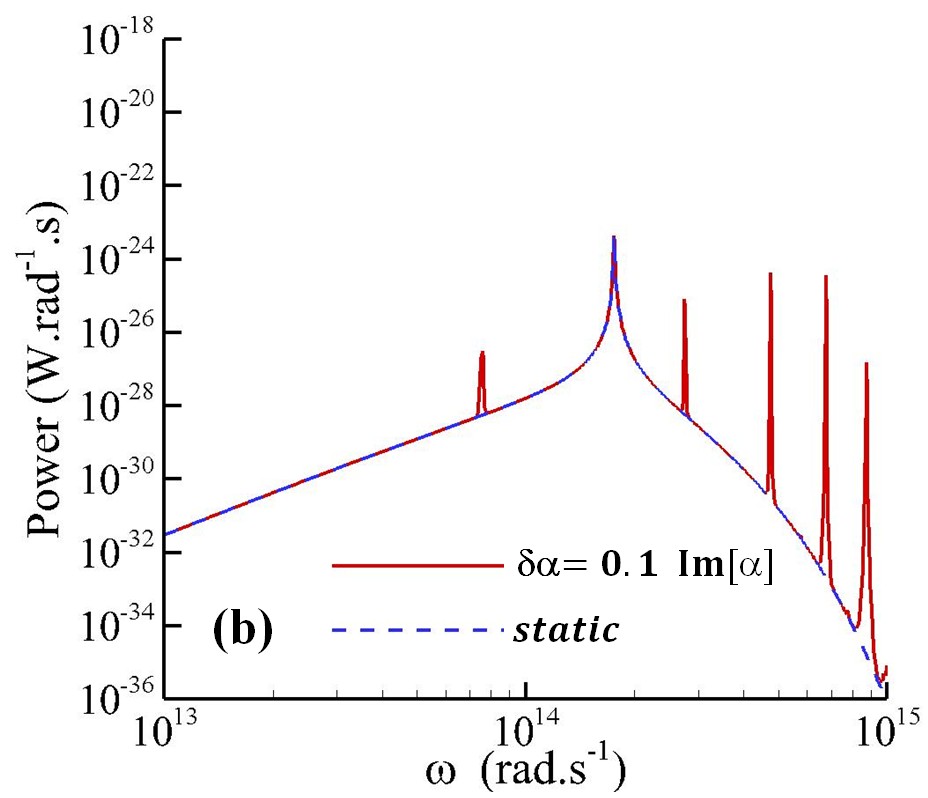}
	\includegraphics[height=0.3\textwidth,angle=0]{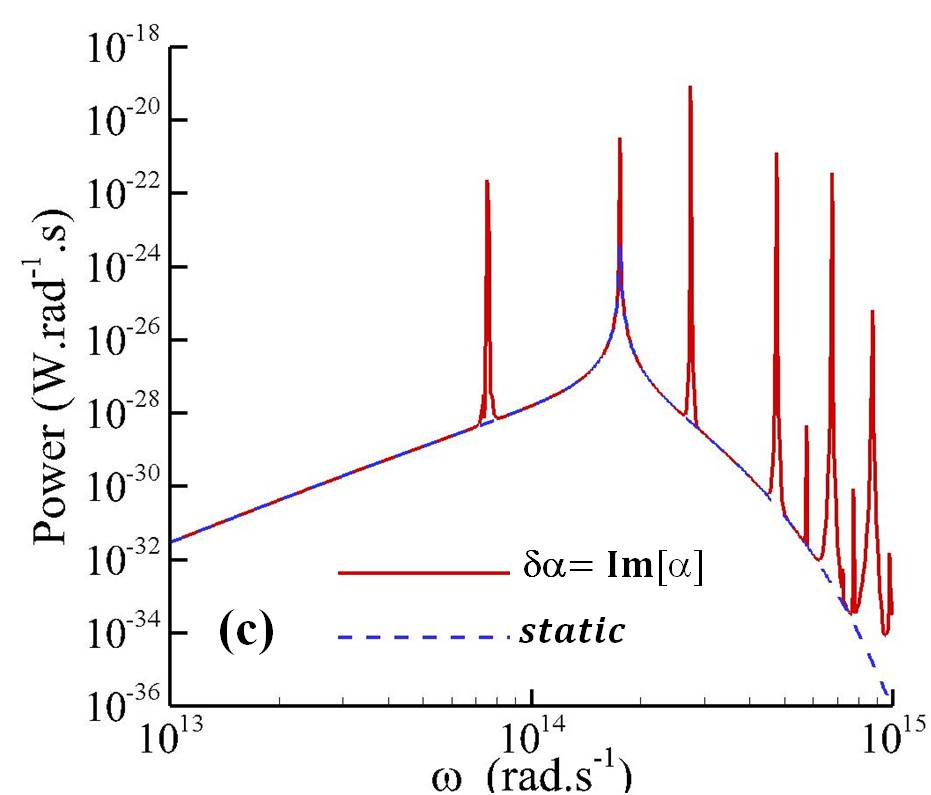}
	\caption{Power spectrum at $T=300\,$K of a modulated dipole (same as in Fig.~\ref{Fig:power_spectrum}) under a rectangular modulation at frequency $\Omega=10^{14}\,\mathrm{rad}\cdot\mathrm{s}^{-1}$ (20 Fourier components ) with various modulation amplitude $\delta\alpha$.}\label{Fig:modulation_amplitude}
\end{figure} 

The thermal emission spectrum in Fig.~\ref{Fig:power_spectrum} is strongly influenced by the interplay between the modulation frequency $\Omega$ and the dipolar resonance frequency $\omega_0$ of the polarizability $\alpha(\omega)$. For a static dipole or slow modulation ($\Omega \ll \omega_0$), the spectrum remains centered near $\omega_0$, with only slight broadening from weak frequency mixing [Figs.~\ref{Fig:power_spectrum}(a),(d)]. When the modulation is much faster than the resonance ($\Omega \gg \omega_0$), the dipole cannot follow the rapid variation of its polarizability. Although frequency mixing persists, the sidebands at $\omega_0 + n\Omega$ lie far from resonance and contribute negligibly, leaving the emission nearly unchanged [Figs.~\ref{Fig:power_spectrum}(c),(f)].

A qualitatively distinct behavior emerges when the modulation frequency approaches the resonance ($\Omega \sim \omega_0$) [Figs.~\ref{Fig:power_spectrum}(b),(e)]. In this near-resonant regime, the dipole acts as a \emph{parametric oscillator}, with the periodic modulation. In that condition, the dipole moment obeys
\begin{equation}
\ddot{p} + \gamma \dot{p} + \omega_0^2 \bigl[1 + \epsilon \cos(\Omega t)\bigr] p = \beta E^{\rm(loc)}(t),
\end{equation}
where $E_{\rm loc}(t)$ is the local driving field. Here, the modulation of the dipole’s polarizability plays the role of a time-dependent spring constant in a mechanical parametric oscillator, while $E^{\rm(loc)}(t)$ acts as an external driving force, illustrating how energy from the modulation can be channeled into selective amplification of thermal fluctuations.

The time-dependent coefficient in both systems couples frequencies separated by integer multiples of $\Omega$, producing sidebands and energy exchange between modes. Near resonance, these sidebands are strongly enhanced, enabling selective amplification of certain spectral components. Physically, the dipole “borrows” energy from the modulation to reinforce thermal fluctuations at sideband frequencies, increasing emission in those regions. Practically, a polar nanosource could be modulated around $10\,$THz by driving its optical phonons with an ultrafast infrared pulse~\cite{Khalsa,Kusaba}, and the resulting time-varying emission spectrum could be detected via time-resolved upconversion~\cite{Cai} or heterodyne infrared~\cite{Paiva} methods. Ultrafast modulation of the dielectric function can also be realized through several distinct mechanisms that produce near–step-like temporal changes in $\varepsilon(t)$. One established route exploits photo‐induced insulator–to–metal transitions in phase-change materials, where femtosecond excitation drives a nonthermal structural and electronic transformation. In $VO_2$, for example, multi-THz spectroscopy has revealed a sub-200-fs collapse of the insulating state and an abrupt rise in metallic conductivity, enabling an almost rectangular modulation of the permittivity~\cite{Pashkin}. A second approach relies on ultrafast optical carrier injection and rapid carrier removal, whereby femtosecond excitation produces sub-cycle changes in the Drude response. Such dynamics, demonstrated in photoconductive THz antennas and related semiconductor systems, yield sub-100-fs variations in the carrier-induced permittivity that appear effectively step-like on THz timescales~\cite{Pedersen,Lu}. A third class of techniques employs intense THz or mid-infrared pumping to drive nonlinear lattice motion in polar crystals. Through anharmonic mode coupling, such nonlinear phononics produces directional ionic displacements and associated shifts in polarization, thereby inducing an ultrafast, non-oscillatory change in the dielectric function~\cite{Först,Handa}.

Regarding the modulation amplitude of the polarizability, it strongly influences the power spectrum, as illustrated in Fig.~\ref{Fig:modulation_amplitude}. For small modulation amplitudes, the impact on the spectrum is approximately quadratic: doubling the amplitude roughly quadruples the contribution from modulation at the sidebands, since the radiated power scales with the square of the dipole moment. Consequently, for weak modulations, the heat flux increases proportionally to the amplitude squared. On the other hand, at larger amplitudes, higher-order effects become significant, leading to a pronounced enhancement of the heat flux.

\section{Conclusion}
In conclusion, we have developed a general framework for radiative heat exchange in dipolar systems with time-modulated polarizabilities, extending fluctuational electrodynamics beyond the stationary regime. Temporal modulation induces memory effects that couple each dipole’s response to its past dynamics, redistributing thermal fluctuations across Floquet sidebands. Our generalized Landauer-like formulation captures all inelastic frequency-conversion channels, revealing how modulation enables selective amplification or suppression of spectral components and providing a unified foundation for tunable and nonreciprocal nanoscale radiative heat transfer.

\end{document}